\documentclass[conference]{IEEEtran}
\IEEEoverridecommandlockouts
\usepackage{cite}
\usepackage{amsmath,amssymb,amsfonts,bm}
\usepackage{algorithmicx}
\usepackage[]{algorithm2e}
\usepackage{graphicx}
\usepackage{textcomp}
\usepackage{xcolor}
\def\BibTeX{{\rm B\kern-.05em{\sc i\kern-.025em b}\kern-.08em
    T\kern-.1667em\lower.7ex\hbox{E}\kern-.125emX}}
\begin{document}

\title{A Generic Algorithm for Sleep-Wake Cycle \\Detection using Unlabeled Actigraphy Data\\
\thanks{This work is supported by NIH UL1TR002649.}
}

\author{\IEEEauthorblockN{Shanshan Chen, \\Robert Perera}\\
\IEEEauthorblockA{\textit{Department of Biostatistics}\\
\textit{Virginia Commonwealth University}\\
Richmond VA, USA \\
\{shanshan.chen, \\robert.perera\}@vcuhealth.org}

\and
\IEEEauthorblockN{Matthew M. Engelhard, \\Jessica R. Lunsford-Avery, \\Scott H. Kollins}
\IEEEauthorblockA{\textit{Department of Psychiatry} \\
\textit{Duke University}\\
Durham NC, USA \\
\{m.engelhard, jessica.r.avery,\\scott.kollins\}@duke.edu}
\and
\IEEEauthorblockN{Bernard F. Fuemmeler}\\ \\
\IEEEauthorblockA{\textit{Department of Health,}\\ \textit{Behavior and Policy} \\
\textit{Virginia Commonwealth University}\\
Richmond VA, USA \\
Bernard.Fuemmeler@vcuhealth.org}
}

\IEEEpubid{\begin{minipage}{\textwidth}\ \\\\\\\\[12pt]
		\copyright 2019 IEEE. Personal use of this material is permitted.  Permission from IEEE must be obtained for all other uses, in any current or future media, including reprinting/republishing this material for advertising or promotional purposes, creating new collective works, for resale or redistribution to servers or lists, or reuse of any copyrighted component of this work in other works.
\end{minipage}}

\maketitle

\begin{abstract}
One key component when analyzing actigraphy data for sleep studies is sleep-wake cycle detection. Most detection algorithms rely on accurate sleep diary labels to generate supervised classifiers, with parameters optimized for a particular dataset. However, once the actigraphy trackers are deployed in the field, labels for training models and validating detection accuracy are often not available.\par
In this paper, we propose a generic, training-free algorithm to detect sleep-wake cycles from minute-by-minute actigraphy. Leveraging a robust nonlinear parametric model, our proposed method refines the detection region by searching for a single change point within bounded regions defined by the parametric model. Challenged by the absence of ground truth labels, we also propose an evaluation metric dedicated to this problem. Tested on week-long actigraphy from 112 children, the results show that the proposed algorithm improves on the baseline model consistently and significantly (p\textless 3e-15). Moreover, focusing on the commonality in human circadian rhythm captured by actigraphy, the proposed method is generic to data collected by various actigraphy trackers, circumventing the laborious label collection step in developing customized classifiers for sleep detection.  
\end{abstract}


\section{Introduction}
Accurate detection of sleep-wake cycles is an essential aspect of sleep research \cite{Meltzer2012}. Wearable activity trackers, such as ActiGraph\textsuperscript{TM} and Fitbit\textsuperscript{\textregistered}, have been widely adopted in sleep research in the last decade. Such trackers usually consist of a 3-axis accelerometer which captures proper acceleration \cite{Sethuramalingam2010} in the forward, lateral, and vertical directions at a sampling rate of up to 100 Hz. When used for tracking daily activity levels, data from the 3-axis accelerometer are often transformed into vector magnitudes and aggregated minute by minute, using various research-based and proprietary algorithms, to give activity counts. \par 
While the promising applications of such activity trackers have excited the fields of sleep research and behavior monitoring, certain challenges ensue in the deployment of such activity trackers. Once the trackers are deployed in the field, labels for training models and validating detection accuracy are often unobtainable, inconvenient to collect, or contain inaccuracies. For example, the customers of commercially-available fitness trackers may not volunteer their sleep diary information, or sleep researchers may elect not to collect such labels in order to minimize the burden on human subjects and maximize the compliance rate. Besides, the accuracy of sleep diary reporting is often unknown, especially when proxy reports are used (e.g. parents reporting for children).\par
The absence of labels in field studies creates two challenges: 1) The development of accurate sleep detection algorithms; and 2) the evaluation of detection accuracy. Most state-of-the-art detection algorithms resort to supervised classifiers, where the activity data are windowed and classified as either ``asleep" or ``awake". The performance of classifiers can then be evaluated using the sleep diary labels and cross-validated using metrics such as accuracy and specificity. As such classifiers are fine-tuned to maximize the detection accuracy for a particular dataset, their generalizability to other datasets is questionable. Without ground truth labels, supervised classifiers cannot be developed.\par
Moreover, regardless of ground truth labels, other issues are associated with the windowing step under a machine-learning framework, using either supervised or unsupervised learning \cite{El2017}. As structural changes in features can happen over various timescales, a fixed window length cannot capture structural changes adequately \cite{Auger1989}. Often, such window-based methods produce fragments of an activity state and the results need to be smoothed by an ad hoc temporal filter (e.g. a median filter) \cite{Cao2012} or heuristic rules, which can further reduce the temporal resolution of the detection.\par

To address these issues, we propose an automated sleep detection algorithm using minute-by-minute unlabeled actigraphy that is generic to various types of wearable activity trackers. Leveraging the prior information obtained from a nonlinear parametric model, we then detect the sleep/wake onset time with higher precision using a parametric change point (CP) detection algorithm. Our contribution lies in the following:
\begin{itemize}
	\item Proposing a generic algorithm that detects sleep/wake cycles in minute-by-minute activity data without sleep diaries or supervised training; 
	\item Proposing an evaluation metric for sleep/wake detection algorithms in the absence of valid ground truth labels.
\end{itemize}

\section{Methods}
\subsection{Data Collection}
The study sample was a subset of children born of mothers who had been recruited for the Newborn Epigenetic Study (NEST) between 2005 and 2011. The women were recruited during pregnancy, and re-contacted  in 2014 when their children were at least three years old. Informed consent was obtained from the mothers for the children to wear ActiGraph\textsuperscript{TM} sensors on their hips to measure their physical activity at home for at least 7 days. After field data collection, the ActiGraph\textsuperscript{TM} sensors were collected by researchers at Duke University for data download and analysis. The study was approved by the Institutional Review Board at Duke University, and the study visits took place at Duke University from 2014 to 2017.

\subsection{Preprocessing}
Among the variables collected by the ActiGraph\textsuperscript{TM} sensor, only the vector magnitude (VM) was used for sleep detection, due to its reliability and pervasiveness. To obtain reliable estimation of circadian rhythms, we excluded subjects who had worn the sensor for less than five days from the analysis. Since we did not require the subjects to wear the sensor at all times, the actigraphy data may contain structural zeros due to non-wearing, mixed with true zeros due to resting, creating problems for sleep detection. In our analysis, we relaxed the 90-minute threshold proposed by Choi et al \cite{Choi2011} by excluding subjects with 120 minutes of consecutive zeros. Using these criteria, we included 112 children's data (aged 7.8$\pm$1.8  years, 52\% female) to develop and test our sleep detection algorithm. 


\subsection{Fitting the Nonlinear Parametric Model}
Since cyclic data can be modeled by fitting periodic curves such as cosine functions, assuming each subject has a single underlying clock, i.e. a constant circadian rhythm, we can fit cosine functions to actigraphy data as well. In fact, Marler et al proposed a sigmoidally-transformed cosine (STC) model to address this problem in 2006 \cite{Marler2006}. By extending a 3-parameter cosine function to a 5-parameter function, the STC model, the solution proposed by Marler et al captured the asymmetries between the awake period (e.g. 16 hours) and the asleep period (e.g 8 hours), as opposed to the equal durations captured by the 3-parameter cosine model described in Equation (1): 
\begin{align}
r(t) = mes + amp \times cos(\dfrac{[t-\phi]\times2\pi}{T})
\end{align}
where t = 1, 2, ..., n, t $\in\mathbb{N}$, is the time index for the actigraphy sequence. T is the period of one's circadian rhythm, often set as a constant of 24 (hours), or in our case 1440 (minutes). Transforming $r(t)$ with the Hill sigmoid function from Equation (2), we can obtain the STC function \cite{Hill1910}:
\begin{align}
h(r) = \dfrac{r^{\gamma}}{m^{\gamma} + r^{\gamma}} 
\end{align}

In Equations (1) and (2), $mes$, $amp$, $\phi$, $\gamma$, and $m$ are the five parameters to be estimated. Interpretations of these parameters in the context of circadian rhythms can be found in \cite{Marler2006}. Here, we are more interested in the segmentation capability implied by the STC curves, especially the parameters related to timing information. To fit the STC model, we used a nonlinear least squares method \cite{Coleman1996} and initialized the parameters with multiple starting points in order to find the global minimum \cite{Ugray2007}. Once we obtained the STC curve, we dichotomized the curve into two states (i.e. ``asleep" or ``awake") by thresholding the STC curve. In our case, we empirically set an adaptive threshold at 20\% of the range of each STC curve, which captured the diurnal activity adequately. The transitions between the two states can be considered as CP detected by STC, denoted by CP\textsubscript{STC}.

\subsection{Change Point Detection}

Consistent sleep/wake times imposed by the STC model do not allow further investigation of sleep irregularity or variability. Therefore, the precise sleep/wake times must be determined with a more refined searching algorithm. Fortunately, once we obtain CP\textsubscript{STC} after fitting the STC model, we have strong prior information about the approximate locations of the true CPs, which should be in the proximity of CP\textsubscript{STC}. Moreover, assuming that one must sleep at least once between two consecutive waking times and wake up at least once between two consecutive sleep onsets, we can identify a single CP between two CP\textsubscript{STC} of the same type. These CPs can be identified using the Pruned Exact Linear Time (PELT) algorithm \cite{Killick2012}, with the actigraphy data modeled as following a gamma distribution \cite{Chen2011}, denoted by CP\textsubscript{PELT}. \par 
\begin{algorithm}
	\caption{Detecting precise sleep/wake onset times guided by STC.}
	\textbf{Input:} $\bm{CP_{STC}} = [CP_{1},CP_{2},...,CP_{m}]$,\\ 
	\quad\quad\quad $\bm{X} = [X_{1},X_{2},...,X_{n}]$; \Comment{time series of actigraphy}\\
	\textbf{Output:} $\bm{CP_{PELT}}$, $\bm{CP_{awake}}$, $\bm{CP_{sleep}}$, $\bm{L}$ \\
	\textbf{Initialize:} set $\bm{CP_{single}}$ to an empty list, \\
	\qquad\quad\quad set k to 0; \Comment{ the number of $\bm{CP\textsubscript{PELT}}$}\\
	Search for the first $CP_{PELT_1}$ in $[X_{1},X_{2},...X_{CP_{2}}]$;\\
	\For {each $CP_{i}$ in $[CP_{2},CP_{3},...,CP_{m-1}]$, } {
		$CP_{Previous}$ = $CP_{PELT_{i-1}}$;\\
		$\bm{M}$ = {$[X_{CP_{Previous}}$, ..., $X_{CP_{i+1}}]$ }\\
        		$\bm{M}$ = $\bm{M}$ + 1e-3; \Comment{shift zeros in M to small positive values such that $M \sim \Gamma(\alpha,\beta)$}\\
		\While{ $\bm{CP_{single}}$ is empty}{
			k = k + 1;\\
			$CP_{single}$ = PELT($\bm{M}$, k, 'gamma');\\
		}
		\If{more than one $CP_{single}$ detected}{
			keep only the $CP_{single}$ closest to $CP_{i}$;\\
		}
		$CP_{PELT}$ = $CP_{Previous}$ + $CP_{single}$;
	}
	Search for the last $CP_{PELT_m}$ in  $[X_{CP_{PELT_{m-1}}}...X_{T}]$;\\
	Identify CP\textsubscript{awake}, CP\textsubscript{sleep} in CP\textsubscript{PELT};\\
	
	Convert  CP\textsubscript{awake} and CP\textsubscript{sleep} to label $\bm{L}$, such that
	$$
	\begin{cases}
	L_{t} = 1, \quad when\quad Awake \\
	L_{t} = 0, \quad when\quad Asleep
	\end{cases}
	$$
\end{algorithm}

Hence, we broke down the problem of searching for multiple CPs in multiple days of actigraphy into searching for a single CP within a well-approximated sleep-wake cycle, which is much easier to solve. Unlike the fixed-length windowing method, the long sequences of multiple days were broken down organically into segments of sleep-wake cycles. The key steps of our proposed algorithm are described in Algorithm 1.\par
With the label vector, $\bm{L}$, obtained from Algorithm 1 and the STC model, we captured diurnal activity and nocturnal activity in two time series vectors $\bm{D}$ and $\bm{N}$:
$$
\begin{cases}
D_{t} = L_{t}\cdot X_{t} ,  \quad\quad\quad t = 1,2,...n \\
N_{t} = (1-L_{t})\cdot X_{t} ,  \quad t = 1,2,...n
\end{cases}
$$

\subsection{Evaluation}
To evaluate the detection performance without ground truth labels, we have to rely on a set of assumptions. First, we assume that nocturnal activity is much less than diurnal activity, thus the ratio between diurnal activity levels and nocturnal activity levels must be much greater than 1. Secondly, underestimating the awake period in $\bm{L}$ will increase the variance of $\bm{N}$; whereas underestimating the sleep period will decrease the variance of $\bm{D}$. Lastly, we consider the STC model as a baseline model which gives the lower bound of detection accuracy. Therefore, any model that improves detection must have larger $R$ than the STC model. Given these assumptions, we define an evaluation metric $R$ as the ratio between the variance of $\bm{D}$ and $\bm{N}$: 
\begin{align}
R = \dfrac{\sum\limits_{t=1}^{n} (D_{t} - \overline{\bm{D}})^2}{ \sum\limits_{t=1}^{n}(N_{t} - \overline{\bm{N}})^2 }
\end{align}
Intuitively, $R$ is sensitive to the detected nocturnal activity. Although a high level of nocturnal activity can also reduce $R$, we argue that diurnal activity mis-identified as nocturnal is likely to be much higher than true nocturnal activity. Moreover, when comparing models, the effect of true nocturnal activity on $R$ can be canceled out. Furthermore, detection results with $R_{Proposed} < \epsilon $ and $R_{Proposed} - R_{STC} < \epsilon $ ($\epsilon$ is a small positive value) can be automatically identified as containing detection errors and re-examined.      

\section{Results}
\begin{figure}[thpb]
	\centering
	\includegraphics[scale=0.35]{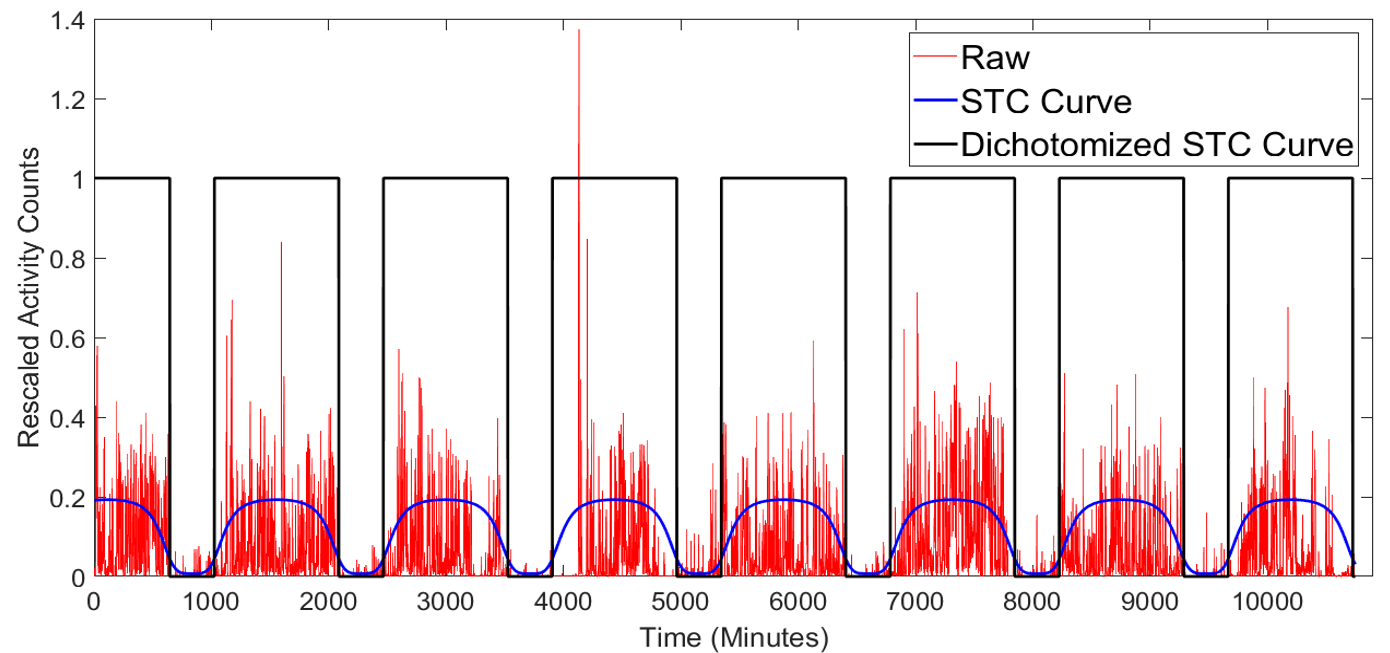}
	\caption{STC model fitting.}
	\label{fig:stc}
\end{figure}

Figure \ref{fig:stc} shows the fitted STC curve and dichotomized STC curve in the form of a binary label vector for an 8-day wearing period. Despite some non-wearing periods in the data, the STC model was able to converge and estimate the underlying circadian rhythm and the transition edges between sleeping and waking.\par  

\begin{figure}[thpb]
	\centering
	\includegraphics[width = 8cm]{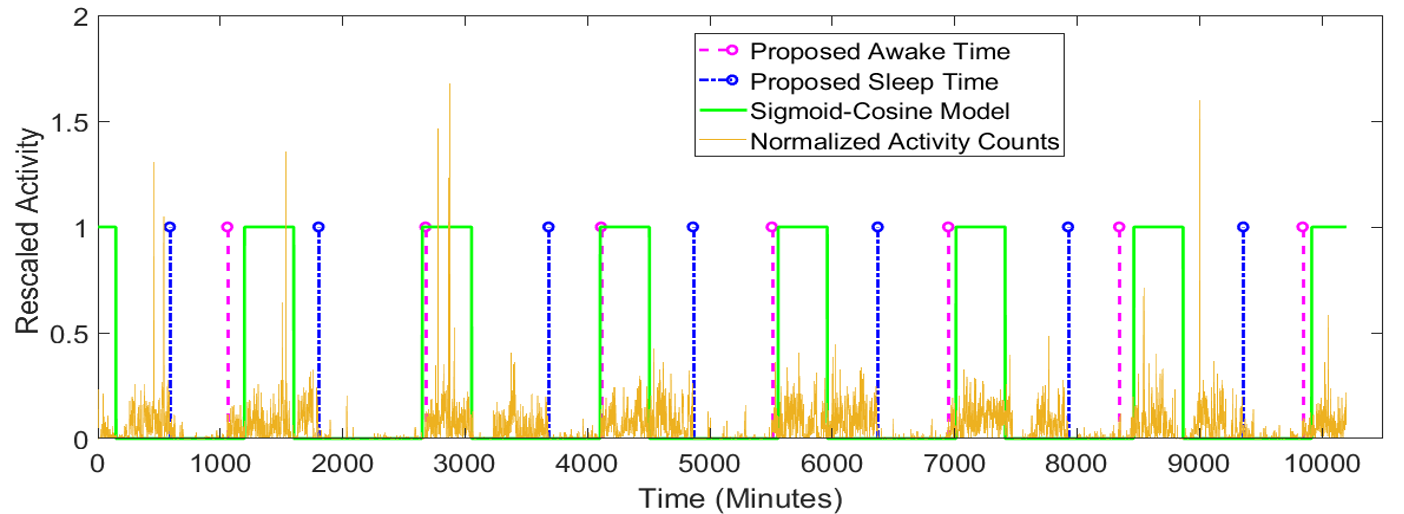}
	\caption{CPs detected by the proposed model improves upon the STC model.}
	\label{fig:cp}
\end{figure}

\begin{figure}[thpb]
	\centering

		\includegraphics[scale=0.35]{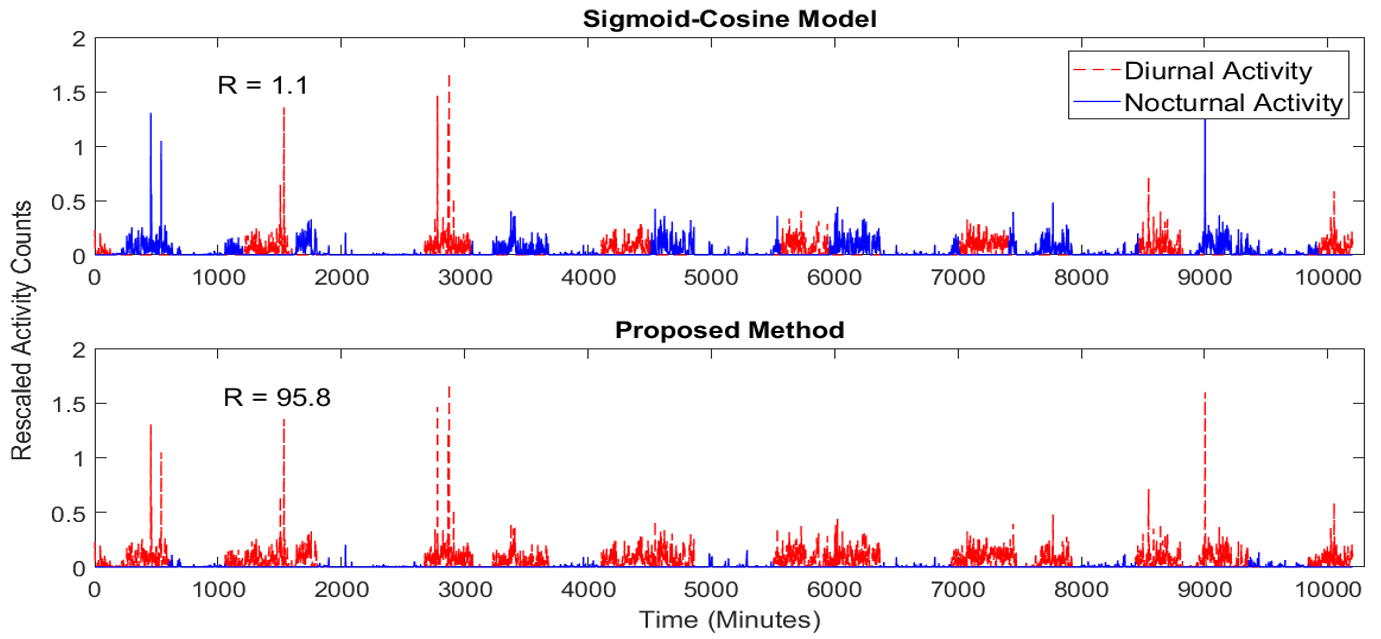}
	\caption{Residual nocturnal activity vs. diurnal activity after applying two types of sleep detection algorithm.}
	\label{fig:residual}
\end{figure}

Figure \ref{fig:cp} shows that the proposed method corrected the transition edges detected by the STC model, in a case where the diurnal activity was severely underestimated by the STC model. Using the same case, Figure \ref{fig:residual} shows the residual diurnal/nocturnal activity after applying the STC model and the proposed method. With the fine-grained search of CP between two transitions, the proposed algorithm corrected the large detection error by presenting much less residual activity in the sleeping period. In this extreme case, $R$ is near 1 for the STC model because the estimated total diurnal activity and total nocturnal activity are nearly the same. Using the proposed model, $R$ is boosted to about 100 by better capturing diurnal activity.

\begin{figure}[thpb]
	\centering
	\includegraphics[scale =0.35]{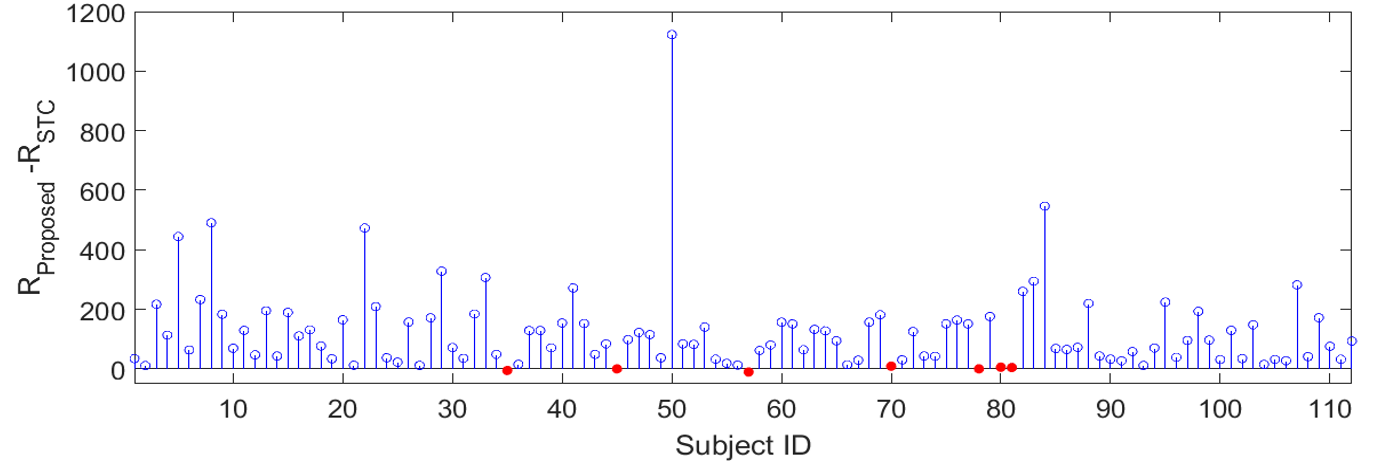}
	\caption{Differences in evaluation metric R.}
	\label{fig:diff}
\end{figure}

Figure \ref{fig:diff} illustrates the difference in detection performance between the STC model and the proposed model evaluated by the proposed metric $R$. Overall, detection results for most subjects significantly improved using the proposed algorithm in comparison with the baseline STC model (p\textless 3e-15, using a one-sided, paired t-test). After visual inspection, we empirically set $\epsilon$ to 10, and found seven subjects' results had small positive or negative values of $R_{Proposed} - R_{STC}$. Researchers can then perform manual inspection on \textit{those} subjects' data, and re-estimate their sleep/wake times, without having to manually inspect \textit{each} subject's data one by one. Overall, with the algorithm implemented in MATLAB 2017a using an Intel Xeon 6-Core CPU at 3.5Hz with 16.0GB RAM, it took 10 $\sim$ 30s to process one subject's actigraphy data.

\section{Discussion}
Our top-down approach converts the sleep/wake cycle detection problem into a change-point detection problem with bounded regions, organically segmented by a nonlinear parametric model. In contrast to previous methods that focused on cutoff points to threshold the activity counts (or metrics derived from them), the assumptions of our method focus on the commonality of sleep patterns in humans. Following the assumption, our method aims to capture the temporal rhythm via a STC model and the day-to-day variability via CP detection bounded by the STC model. Therefore, although our data was collected on children using a hip-worn sensor, the method is generalizable to other types of actigraphy, given visually apparent circadian rhythms and reasonably modeled data distributions.\par
In the entire process, the only parameter that needs to be chosen in advance is the threshold to dichotomize the STC curve. Although we used a set of priors to initialize the STC model, those values are related to the circadian rhythms of human beings, thus can be applied to other datasets without tuning. Moreover, with the well-bounded sleep-wake cycles detected, this generic algorithm can also be extended hierarchically to detect naps, disturbed sleep episodes, and sleep stages. \par
One limitation of our evaluation metric is the assumption that activity during the true sleep state is usually lower than activity while awake. However, for subjects with severely disturbed sleep, this may not be the case. Still, this intuitive evaluation metric does not rely on any sleep diary labels or polysomnography data, and is sensitive to sleep/wake detection errors, providing an automatic screening method for wrongly-detected sleep/wake cycles. Another disadvantage of the proposed algorithm is that the computation time can be slightly longer than directly applying discriminative classifiers when finding globally optimized parameters to fit the STC model. Nevertheless, since the proposed algorithm circumvents the laborious collection of sleep labels and training classifiers for various actigraphy datasets (collected by different devices with different configurations), this automated approach reduces the overall overhead in developing sleep detection algorithms. 

\section{CONCLUSIONS}
We proposed a generic, training-free algorithm to detect sleep/wake onset times using field actigraphy data absent of ground truth labels. We also proposed an intuitive metric to evaluate the performance of our detection algorithm and compared it with a nonlinear parametric model. The results show that our proposed detection algorithm improves detection precision, whilst maintaining the accuracy imposed by the parametric model. Work is underway to validate the method and the effectiveness of the proposed metric with ground truth polysomnography data, and to apply the model to other actigraphy datasets. 

\addtolength{\textheight}{-12cm}   

\bibliographystyle{IEEEtran}

\bibliography{Generic_Sleep}

\begin{thebibliography}{10}
\providecommand{\url}[1]{#1}
\csname url@rmstyle\endcsname
\providecommand{\newblock}{\relax}
\providecommand{\bibinfo}[2]{#2}
\providecommand\BIBentrySTDinterwordspacing{\spaceskip=0pt\relax}
\providecommand\BIBentryALTinterwordstretchfactor{4}
\providecommand\BIBentryALTinterwordspacing{\spaceskip=\fontdimen2\font plus
\BIBentryALTinterwordstretchfactor\fontdimen3\font minus
  \fontdimen4\font\relax}
\providecommand\BIBforeignlanguage[2]{{%
\expandafter\ifx\csname l@#1\endcsname\relax
\typeout{** WARNING: IEEEtran.bst: No hyphenation pattern has been}%
\typeout{** loaded for the language `#1'. Using the pattern for}%
\typeout{** the default language instead.}%
\else
\language=\csname l@#1\endcsname
\fi
#2}}

\bibitem{Meltzer2012}
L.~J. Meltzer, H.~E. Montgomery-Downs, S.~P. Insana, and C.~M. Walsh, ``Use of
  actigraphy for assessment in pediatric sleep research,'' \emph{Sleep medicine
  reviews}, vol.~16, no.~5, pp. 463--475, 2012.

\bibitem{Sethuramalingam2010}
T.~Sethuramalingam and A.~Vimalajuliet, ``Design of mems based capacitive
  accelerometer,'' in \emph{Mechanical and Electrical Technology (ICMET), 2010
  2nd International Conference on}.\hskip 1em plus 0.5em minus 0.4em\relax
  IEEE, 2010, pp. 565--568.

\bibitem{El2017}
Y.~El-Manzalawy, O.~Buxton, and V.~Honavar, ``Sleep/wake state prediction and
  sleep parameter estimation using unsupervised classification via
  clustering,'' in \emph{2017 IEEE International Conference on Bioinformatics
  and Biomedicine (BIBM)}.\hskip 1em plus 0.5em minus 0.4em\relax IEEE, 2017,
  pp. 718--723.

\bibitem{Auger1989}
I.~E. Auger and C.~E. Lawrence, ``Algorithms for the optimal identification of
  segment neighborhoods,'' \emph{Bulletin of mathematical biology}, vol.~51,
  no.~1, pp. 39--54, 1989.

\bibitem{Cao2012}
H.~Cao, M.~N. Nguyen, C.~Phua, S.~Krishnaswamy, and X.~Li, ``An integrated
  framework for human activity classification.'' in \emph{UbiComp}, 2012, pp.
  331--340.

\bibitem{Choi2011}
L.~Choi, Z.~Liu, C.~E. Matthews, and M.~S. Buchowski, ``Validation of
  accelerometer wear and nonwear time classification algorithm,''
  \emph{Medicine and science in sports and exercise}, vol.~43, no.~2, p. 357,
  2011.

\bibitem{Marler2006}
M.~R. Marler, P.~Gehrman, J.~L. Martin, and S.~Ancoli-Israel, ``The sigmoidally
  transformed cosine curve: a mathematical model for circadian rhythms with
  symmetric non-sinusoidal shapes,'' \emph{Statistics in medicine}, vol.~25,
  no.~22, pp. 3893--3904, 2006.

\bibitem{Hill1910}
A.~V. Hill, ``The possible effects of the aggregation of the molecules of
  haemoglobin on its dissociation curves,'' \emph{j. physiol.}, vol.~40, pp.
  4--7, 1910.

\bibitem{Coleman1996}
T.~F. Coleman and Y.~Li, ``An interior trust region approach for nonlinear
  minimization subject to bounds,'' \emph{SIAM Journal on optimization},
  vol.~6, no.~2, pp. 418--445, 1996.

\bibitem{Ugray2007}
Z.~Ugray, L.~Lasdon, J.~Plummer, F.~Glover, J.~Kelly, and R.~Mart{\'\i},
  ``Scatter search and local nlp solvers: A multistart framework for global
  optimization,'' \emph{INFORMS Journal on Computing}, vol.~19, no.~3, pp.
  328--340, 2007.

\bibitem{Killick2012}
R.~Killick, P.~Fearnhead, and I.~A. Eckley, ``Optimal detection of changepoints
  with a linear computational cost,'' \emph{Journal of the American Statistical
  Association}, vol. 107, no. 500, pp. 1590--1598, 2012.

\bibitem{Chen2011}
J.~Chen and A.~K. Gupta, \emph{Parametric statistical change point analysis:
  with applications to genetics, medicine, and finance}.\hskip 1em plus 0.5em
  minus 0.4em\relax Springer Science \& Business Media, 2011.

\end{thebibliography}

\end{document}